\documentclass[aps, prl, superscriptaddress, reprint]{revtex4-2}
\usepackage{amsmath}
\usepackage{bbold}
\usepackage{mathdots}
\usepackage{color}
\usepackage{bm,graphicx,hyperref}

\hypersetup{%
  breaklinks = {true},
  citecolor = {blue},
  colorlinks = {true},
  linkcolor = {red},
}

\begin{document}

\title{Microscopic theory of ionic motion in solid electrolytes}

\author{Aleksandr Rodin}
\affiliation{Yale-NUS College, 16 College Avenue West, 138527, Singapore}
\affiliation{Centre for Advanced 2D Materials, National University of Singapore, 117546}

\author{Keian Noori}
\affiliation{Centre for Advanced 2D Materials, National University of Singapore, 117546}

\author{Alexandra Carvalho}
\affiliation{Centre for Advanced 2D Materials, National University of Singapore, 117546}

\author{and A. H. Castro Neto}
\affiliation{Centre for Advanced 2D Materials, National University of Singapore, 117546}
\affiliation{Department of Materials Science Engineering, National University of Singapore, 117575}

\date{\today}

\begin{abstract}

We propose a microscopic, first-principles description of the ionic conduction in crystals. This formalism allows us to gain new insights into the ideal characteristics of general ionic conducting materials and, in particular, solid electrolytes. Using \textit{ab initio} calculations, we show that our formalism results in ionic mobilities consistent with experiments for several materials.  Our work opens the possibility of developing solid electrolytes based on fundamental physical principles rather than empirical descriptions of the underlying processes.  
\end{abstract}	

\maketitle

The semiconductor technology revolution that occurred in the middle of the 20th century has its origin in the development of the microscopic theory of electron motion in crystals in the early days of quantum mechanics. Trailblazers such as F. Bloch showed the importance of taking into account the periodicity of the crystal structure in order to understand electronic behavior and the properties of many different types of materials, from metals to semiconductors~\citep{Hoddeson1987}. Not only did this theoretical framework become important for the understanding of naturally occurring materials, it also allowed for the development of new materials with tailored properties that did not exist before. Electronics turned from a heuristic discipline, based on trial and error, into a predictive science and technology. With the development of superior semiconductor-based technologies, such as the one based on complementary metal–oxide–semiconductors (CMOS), these basic concepts became a common language between scientists and engineers. Such extraordinary developments were driven by the necessity of replacing the obsolete vacuum tube technology that was ubiquitous to the electronic devices of that era.

Arguably, we find ourselves in a similar situation in the area of energy storage. The electrochemical science that has been the basis of battery technology for the last 200 years is faced with the necessity of reinventing itself in order to fulfil the societal needs of today, namely, safe, efficient, reliable, fast, and long-lasting energy storage devices that can be seamlessly incorporated into a modern environmentally conscious lifestyle~\citep{IEA2020,Yang2018,Wang2019}. The pressure felt by important industrial sectors, such as the automotive, has inspired scientists and engineers to look for alternative solutions to traditional approaches. As such, the field of solid-state batteries has emerged as a possible solution to the conundrum of developing commercially viable electric vehicles\citep{Bachman2016,Manthiram2017,Famprikis2019}.

In a solid-state battery, the movable ions, such as lithium ($\mathrm{Li}^+$), traverse a crystal, that is, a periodic structure consisting of fixed ions (that we call the framework) along with their electrons, and interact with these elements via strong Coulomb forces. Unlike the traditional electrochemical problem in a liquid medium, the ionic motion in crystals depends strongly on their periodicity and symmetries, as in the case of Bloch’s theorem. The problem at hand is akin to the famous many-body problem found in strongly interacting electron materials (where phenomena such as magnetism and superconductivity occur) with some fundamental differences, namely, the mass of an ion is approximately 10,000 times that of an electron and while electrons have a strong wave-like characteristics, ions are essentially particles with atomic size. The intricate dance between particles and waves in a crystalline environment is what determines the ionic conductivity and the ultimate efficiency of a solid electrolyte in a battery.

The objective of this work is to develop the basic principles and microscopic formalism that describe ionic motion in crystalline electrolytes. Instead of looking at the problem from a traditional electrochemical perspective, we take a modern condensed matter approach and include the basic elements (mobile and fixed ions and their electrons, in addition to the periodicity of the crystal) from the very beginning. We obtain the steady state equation for the ion motion in a crystal and show that it obeys a Langevin dynamics, consistent with the fluctuation-dissipation theorem, where the ion mobility is determined by the curvature profile of the potential in the atomic lattice and the low frequency phonons of the framework.

In order to substantiate our results, we make use of \textit{ab initio} density functional theory (DFT) to calculate the ion mobility for some crystalline electrolytes and show that our results are consistent with experimental values.

The microscopic Hamiltonian for the problem is given by:
\begin{equation}
    H = K_{e} + V_{ee} + K_{i} + V_{ii} + V_{ei},
    \label{eqn:microscopic_hamiltonian}
\end{equation}
where $K$ refers to the kinetic energy and $V$ to the Coulomb potential of interactions for electrons (subscript $e$) and ions (subscript $i$). Given that the mass of the ions is much larger than that of the electrons, we can consider mobile and framework ions to be static in the time scale of motion of electrons (the Born-Oppenheimer approximation). Furthermore, a solid-state electrolyte is characterized to be an insulator for electrons, which implies that the material has a large band gap in the electronic structure so that electron-hole excitations are not created during the ionic motion. Consequently, one can use the adiabatic theorem and assume that the ionic motion only leads to smooth modifications of the electron-ion interaction. Within these two standard approximations one is left with an effective ion-ion interaction, $U(\mathbf{r}, \mathbf{u})$, which depends on the position of the mobile ions, $\mathbf{r}$, and the framework ions, $\mathbf{u}$ (mathematical details will be presented
elsewhere).

\begin{figure}
    \centering
    \includegraphics[width=8cm]{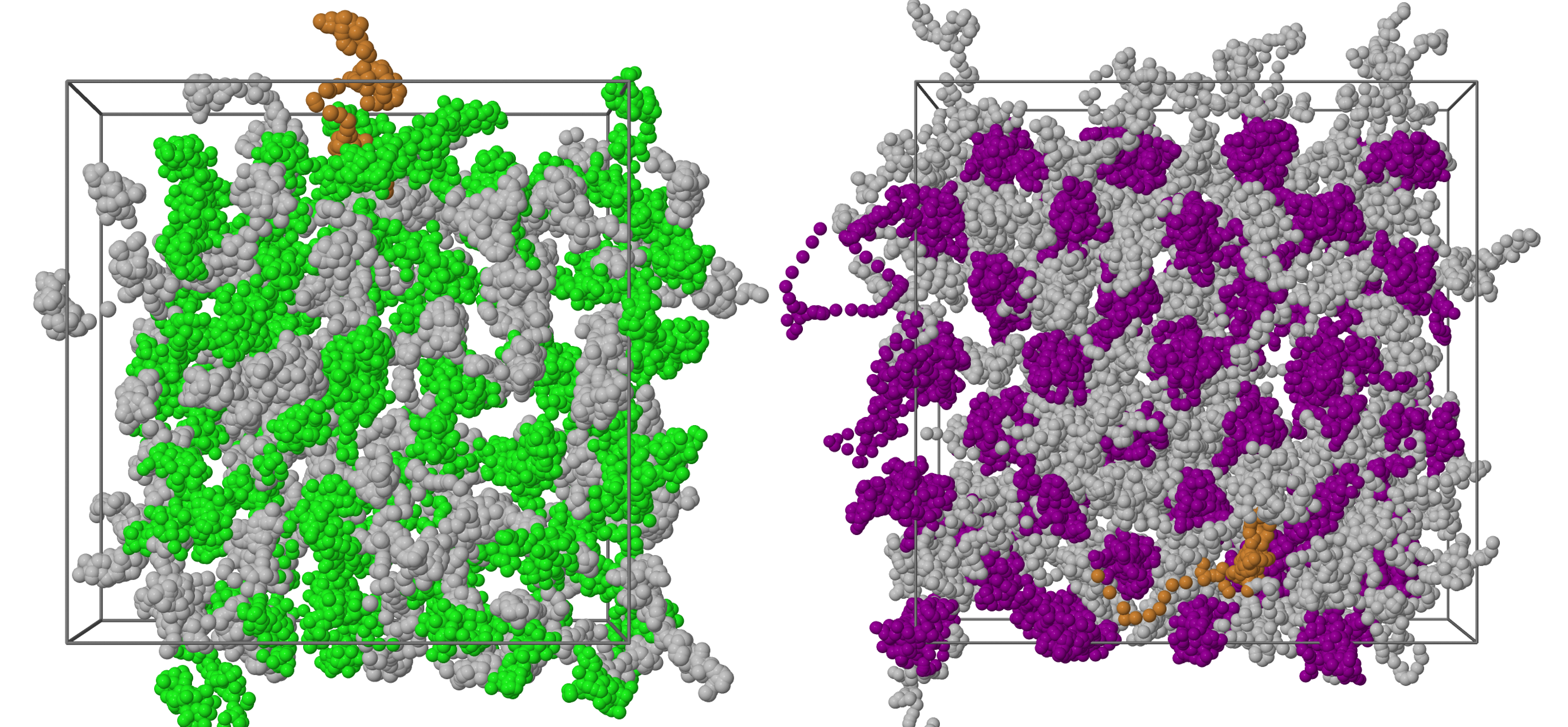}
    \caption{Diffusion trajectories of ions in AgCl at 600 K (left panel) and $\alpha$-AgI at 700~K (right panel), obtained from \textit{ab initio} molecular dynamics simulations.  Ag atoms are represented in grey, while Cl and I atoms are represented in green and purple, respectively. The trajectory of a single Ag ion is highlighted in orange. The positions are represented every 0.1~ps, for a total time of 10~ps.}
    \label{fig:trajectories}
\end{figure}

By definition, framework ions are the ones that remain in their crystal lattice position during the flow of the mobile ions. Hence, at any temperature $T$ below the melting point of the crystal, the framework ions undergo oscillatory motion around their equilibrium positions, as illustrated in Fig.~\ref{fig:trajectories}.

Once again, we can use standard solid-state language to describe the framework ions in terms of their phonons, which are characterized by their frequencies $\Omega_{s}$, where the subscript $s$ labels the phonon modes. Of particular importance are the acoustic phonons, or sound modes, with speed of propagation $v_{L,T}$ (where $L$ and $T$ labels the longitudinal and transverse modes). One should stress that these modes do not exist in a liquid.  They only exist in a crystal because a periodicity is induced by the presence of the lattice.  

Further, the diffusion of ions in solids is different from Brownian motion in a liquid, where the potential $U({\bf r})$ is constant. To illustrate this point, consider mobile ions travelling through a solid. These could be excess ions,  for example in AgI that is Ag-rich due to I evaporation~\cite{fletcher1971reply}, or be thermally generated, as for example the interstitial moieties of Frenkel pairs in AgCl~\cite{friauf1977determination}. These mobile ions travelling through the solid have to regularly escape local potential minima assisted by the framework's thermal fluctuations. This motion resembles a ``hopping" transport, where the ions oscillate around a local minimum before moving to an adjacent one. This is evident in the trajectories obtained from molecular dynamics simulations of the thermal diffusion in the ionic conductors AgCl and $\alpha$-AgI (Fig.~\ref{fig:trajectories}). In both AgCl, where individual Ag ions sporadically escape deep local energy minima, leaving a vacancy behind, and in $\alpha$-AgI, where the Ag sublattice is nearly melted, the mobile ions do not move in straight lines; instead, their trajectories avoid other ions to minimize repulsion. This is the case even in the ``molten” sublattice of superionic conductors, of which the Ag sublattice of $\alpha$-AgI is a typical example.

In the hopping regime, the idea of ``drag" is not quite applicable. Rather, in  superionic conductors the mobile ions flow through the framework in a quasi-free manner, being deflected without becoming trapped by the framework atoms.
Additionally, applying a constant external electric field leads to a dissipative current.

We use the non-equilibrium Keldysh formalism in the path integral representation in order to get the equation of motion for the mobile ions in the form of Newton’s equation: 
\begin{equation}
    M\ddot{\mathbf{r}}(t) = -\nabla U(\mathbf{r},\mathbf{u}_0)-\gamma(\mathbf{r})\dot{\mathbf{r}}(t) + \tilde{\mathbf{f}}(t) + \mathbf{F}(\mathbf{r})\,,
    \label{eqn:EOM}
\end{equation}
where $t$ is time, $\mathbf{u}_0$ is the equilibrium position of the framework ions ($\nabla$ is the gradient with respect to $\mathbf{r}$) $M$ is the ion mass, $\mathbf{F}$ is an applied external force (i.e., electric field), 
\begin{equation}
    \gamma(\mathbf{r}) = 2\pi \sum_s \nabla Y_s(\mathbf{r})\otimes \nabla Y_s(\mathbf{r}) \frac{\delta(\Omega_s)}{\hbar\Omega_s}
    \label{eqn:gamma}
\end{equation}
is the dissipative tensor [$\delta(\mathbf{r})$ is the Dirac delta function], where 

\begin{equation}
    Y_s(\mathbf{r}) =
	\sqrt{\frac{\hbar}{2\Omega_s}}
	\left[\nabla_{\mathbf{u}^0} U\left(\hat{\mathbf{r}},\mathbf{u}^0\right)
    \right]^T
    \mathbf{m}^{-1/2}
	\boldsymbol{\varepsilon}_{s}
	\label{eqn:Y}
\end{equation}
($\hbar$ is Planck’s constant, $\mathbf{m} = \bigoplus_j m_j \mathbf{1}_{D\times D}$ is the framework mass matrix for ions with mass $m_j$ and $D$ is the system dimensionality, $\varepsilon_s$ is the phonon polarization vector). In Eq.~\eqref{eqn:EOM}, the first term on the r.h.s. is a classical term that describes the periodic potential of the static lattice with the framework ions fixed to their equilibrium positions. The second term on the r.h.s. describes the dissipation of energy due to the ion motion when the mobile ion interacts with the phonons of the framework ions. We note that unlike the dynamics in liquids, the dissipative term is position- and direction-dependent, and described in terms of a tensor. Hence, dissipation in a crystal is not isotropic as in a fluid and has quantum nature (as can be seen from the presence of Planck’s constant in its definition). This term is rather non-trivial, as we will see below. $\tilde{\mathbf{f}}(\mathbf{r})$  is the fluctuation force due to the vibrations of the framework. Its origin is the same as the dissipative term. One can show that its correlation function is given by:
\begin{align}
    \langle \tilde{\mathbf{f}}(t)\otimes\tilde{\mathbf{f}}(t')\rangle
	=
	\sum_s 
     &\nabla Y_s\left[\mathbf{r}(t)\right]
     \otimes
     \nabla Y_s\left[\mathbf{r}(t')\right]
     \nonumber
     \\
     \times&
     \coth\left(\frac{\beta\hbar\Omega_s}{2}\right)\cos\left[ \Omega_s\left(t - t'\right)\right]\,,
     \label{eqn:noise_correlation}
\end{align}
where $\beta^{-1} = k_BT$ and $k_B$ is the Boltzmann constant. At high temperatures, $k_BT \gg \hbar \Omega_s$, we can readily see that:

\begin{equation}
    \gamma(t) = \frac{1}{k_BT}\int_{-\infty}^tdt'\langle \tilde{\mathbf{f}}(t)\otimes\tilde{\mathbf{f}}(t')\rangle\,,
    \label{eqn:gamma_high_T}
\end{equation}
which is a generalized fluctuation-theorem for ionic motion in crystals. Hence, dissipation and fluctuation are intimately related. Once again, unlike in liquids, the fluctuation forces are position- and direction-dependent. 

Having established the existence of a Langevin dynamics of mobile ions in a crystal lattice, we can simplify the problem further by considering the steady-state motion in the presence of a constant applied electric field, $\mathbf{E}$. In this case, the acceleration and fluctuation terms vanish, and Eq.~\eqref{eqn:EOM} can be solved for the ion velocity, $\mathbf{v}(\mathbf{r}) = \dot{\mathbf{r}}(t)$:

\begin{equation}
    \mathbf{v}(\mathbf{r}) = \gamma^{-1}(\mathbf{r})\left[-\nabla U(\mathbf{r}) + q \mathbf{E}\right]\,,
    \label{eqn:v_r}
\end{equation}
where $q$ is the ion charge. We now can use the fact that the lattice potential and the dissipation have the periodicity of the lattice and, hence, can be expanded in a Fourier series in terms of reciprocal lattice vectors, $\mathbf{K}$, in order to get:

\begin{equation}
    \mathbf{v}_\mathbf{K} = (2\pi)^{3/2}\sum_{\mathbf{K}'}\gamma^{-1}_{\mathbf{K}-\mathbf{K}'}\left[-i\mathbf{K}'U_{\mathbf{K}'}+q \mathbf{E} \delta_{\mathbf{K}',0}\right]
    \label{eqn:v_K}
\end{equation}
which can be thought of as the equivalent of Bloch’s theorem for ionic motion in a crystal. If we are interested only in the average drift velocity we can take $\mathbf{K}\rightarrow 0$ in the above equation and find:

\begin{equation}
    \mathbf{v}_\mathrm{drift} = \mathbf{v}_{\mathbf{K}\rightarrow 0}=\langle \gamma^{-1}\rangle q\mathbf{E}\,, 
    \label{eqn:v_drift}
\end{equation}
where

\begin{equation}
    \langle \gamma^{-1}\rangle = (2\pi)^{3/2}\sum_{\mathbf{K}}\gamma_\mathbf{K}^{-1}\,,
    \label{eqn:avg_gamma}
\end{equation}
is the lattice-averaged dissipation coefficient. From the above expression we can readily obtain the ion mobility:

\begin{equation}
    \mu = \frac{ \mathbf{v}_\mathrm{drift}}{E} = q\langle \gamma^{-1}\rangle\,,
    \label{eqn:mu}
\end{equation}
where $\langle \gamma^{-1}\rangle$ can be computed from first principles for any crystal lattice. 

In order to gain more insight into the solution, we can simplify the problem considerably by assuming that in Eq.~\eqref{eqn:noise_correlation} only the acoustic phonon modes contribute to the mobility. In this case one can show that:

\begin{equation}
    \gamma(\mathbf{r}) = \frac{1}{12\pi \rho}\left(\frac{1}{v_L^3}+\frac{2}{v_T^3}\right)\left[\mathbf{H}_\mathbf{r}U(\mathbf{r})\right]^2
    \label{eqn:gamma_Hessian}
\end{equation}
where $\rho$ is the mass density of the crystal and $\mathbf{H}_\mathbf{r}$ is the Hessian operator. We now can see very clearly the dissipative mechanism of ionic motion in a crystal, namely, the softer the crystal (smaller sound velocity) the more dissipative the ionic motion. Furthermore, the movement of the ion is dissipationless in regions of the potential where the Hessian vanishes, which are the saddle point regions of the periodic potential in the unit cell of the crystal. Finally, this expression gives us clues regarding what kinds of crystals would be good electrolytes, namely, hard crystals with smooth potential configurations.  

\begin{figure}
    \centering
    \includegraphics{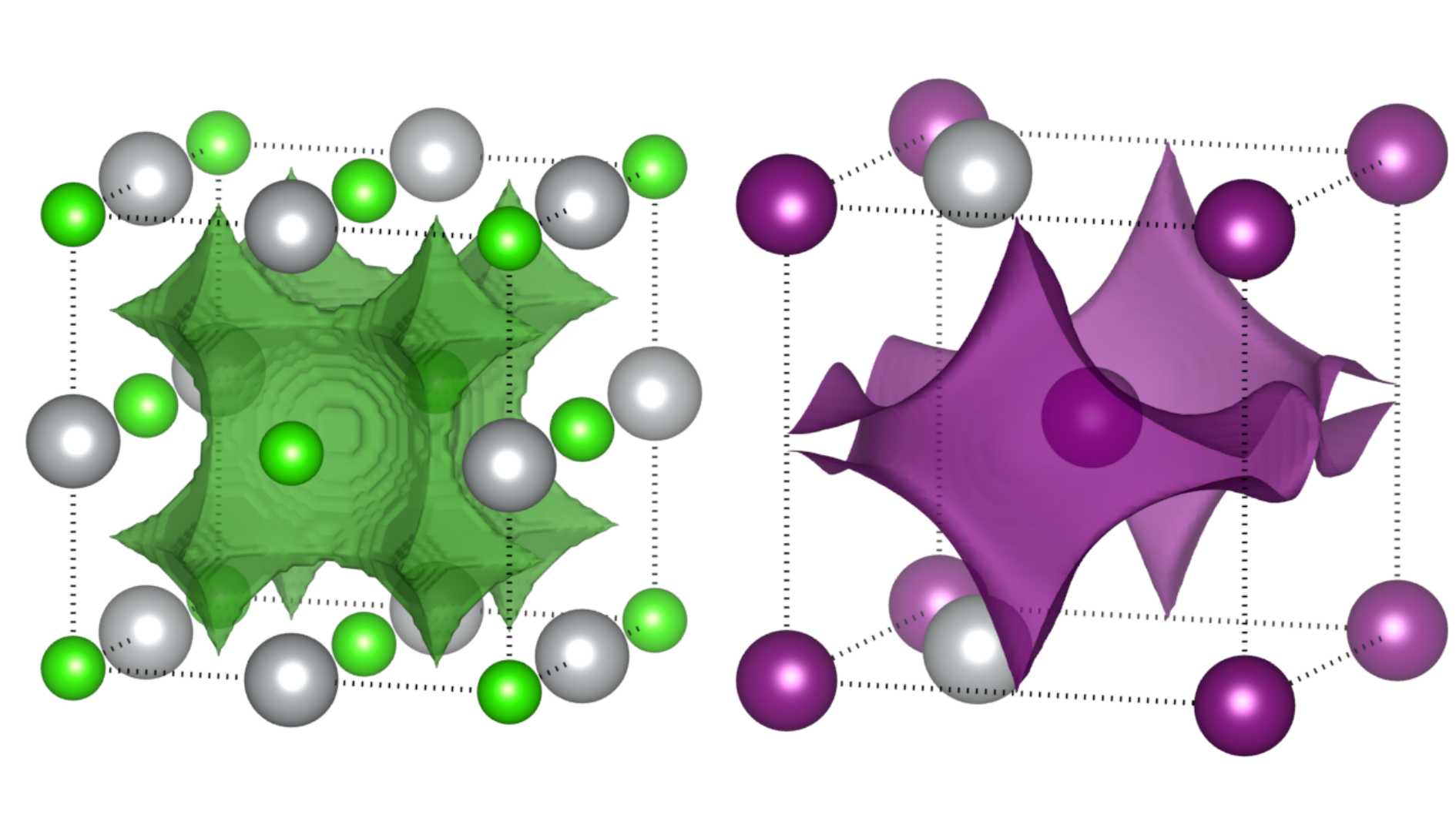}
    \caption{Three-dimensional potential energy profile, $U(\textbf{r})$, of a mobile Ag ion for AgCl (left panel) and $\alpha$-AgI (right panel). The isosurfaces show the minimum energy at which a continuous connecting pathway exists for the unit cell. The positions of the fixed ions are indicated. Ag atoms are represented in gray, while Cl and I atoms are represented in green and purple, respectively.}
    \label{fig:potential_map}
\end{figure}

The variations of the potential energy surface can be quantified by computing $U(\mathbf{r})$ from first principles, which we do, as illustration, for the metal-halide electrolytes AgCl, LiCl, LiI, $\alpha$-AgI, and $\alpha$-CuBr, as shown in Fig.~\ref{fig:potential_map} for AgCl and $\alpha$-AgI. From here, we can obtain $\mu$, per mobile ion, for each compound via the calculation of $\gamma(r)$, as defined in Eq.~\eqref{eqn:gamma_Hessian}. These ion mobilities, $\mu_{calc}$, assuming $q=e$, are listed in Table~\ref{tab:mobilities}. A precise comparison between experiment and theory is made difficult by the presence of non-idealities in experimental samples; for instance, experimental samples are often polycrystalline, may exhibit size effects, and may have more than one mobile defect or ionic species. In view of these complexities, the consistency between the calculated and experimentally-extracted mobilities is notable, especially when considering the simple nature of the \textit{ab initio} ingredients involved. Moreover, the accuracy of the calculated mobilities can be readily improved by enforcing more stringent criteria on, e.g., the smoothness of the energy isosurface or the size of the supercell.

\begin{table}[]
    \centering
    \begin{ruledtabular}
    \begin{tabular}{cccccc}
     compound & AgCl & LiCl & LiI & $\alpha$-AgI & $\alpha$-CuBr \\
     \hline
     $\mu_{calc}$ ($\mathrm{cm^{2}/V~s}$)  & 0.0012 & 0.043 & 0.99 & 0.25 & 0.18 \\
     $\mu_{expt}$ ($\mathrm{cm^{2}/V~s}$) & 0.08 & - & 0.13 & 0.0022 & $\simeq$ $10^{-3}$ \\
     $E_{a,expt}$ (eV) & 0.71 & - & 0.42 & 0.054 & - \\
    \end{tabular}
    \end{ruledtabular}
    \caption{Calculated ionic mobilities per mobile ion ($\mu_{calc}$), together with the values obtained from fitting previous experimental reports ($\mu_{expt}$). The calculated values are directionally-averaged ionic mobilities computed via $\gamma$ (Eq.~\eqref{eqn:gamma_Hessian}), using the potential from \textit{ab initio} calculations. The experimental values for the ion mobility and for the activation energy for creation of a mobile ion were obtained from fitting experimental data (see Methods section for details) - AgCl: from Ref.~\citenum{Maier1988}; LiI: from Ref.~\citenum{Poulsen1980}; $\alpha$-AgI: from Ref.~\citenum{Sunandana2004}; $\alpha$-CuBr: from Ref.~\citenum{Funke2013}.}
    \label{tab:mobilities}
\end{table}

In summary, we have developed a microscopic theory for ionic motion in crystals and derived the equivalent of Bloch’s theorem for ions. We found that the ionic mobility depends essentially on the lattice softness (via the third power of the sound velocity) and the curvature of the atomic potential felt by the ions; namely, hard materials with smooth atomic potentials are the best candidates for high ionic mobility. This theory yields a tractable route for the calculation of ionic mobilities via modern \textit{ab initio} or other theoretical methods.  Further, the \textit{ab initio} approach can, in principle, be extended to account for the extrinsic effects that impact measured ionic mobilities, such as grain boundaries, impurities, and other types of defect that are already well-known in solid-state physics. Our calculations for the ionic mobility of several ionic conductors using \textit{ab initio} methods provide an upper bound for the mobility in single crystals and indicate that the currently measured ionic motilities found in the literature are dominated by extrinsic effects such as impurities, grain boundaries, and other types of defects that are already well-known in solid-state physics. Although in the last century we have developed a powerful theoretical framework to study the effect of defects and interfaces in the motion of electrons in solids, the same cannot be said for the case of ions.  The understanding of how ions interact with defects and interfaces in solids is an unexplored land. Any further progress in the development of solid-state electrolytes, which are the key elements of solid-state batteries, depends fundamentally on progress in this area of research.   

\section*{Methods}
\paragraph{Density functional theory at 0~K --}
DFT calculations are performed using the Quantum \textsc{ESPRESSO}~\citep{Giannozzi2009,Giannozzi2017} code. Structural relaxations and total energy calculations are performed using a PAW basis~\citep{Blochl1994,DalCorso2014} and the Perdew-Burke-Ernzerhof (PBE) exchange-correlation functional~\citep{Perdew1996}. The kinetic energy cutoffs of the charge density and wavefunctions are set to at least the minimum recommended values of the PAW pseudopotential~\citep{DalCorso2014}. The Brillouin zones for all materials are sampled using uniform grids of $4\times4\times4$ (total energies) and $6\times6\times6$  (phonons) K-points.  

For the calculation of $U(r)$ we allow one ion of the mobile metal species to move while keeping all other ions fixed. The mobile ion is moved within the cubic unit cell by intervals of $1/64$ of the lattice parameter, $a$. Only configurations in which the distance from the mobile ion to any fixed ion is greater than ($5/12$)$a$ ($\alpha$-AgI and $\alpha$-CuBr) or ($1/3$)$a$ (AgCl, LiCl, LiI) are permitted. In the cases of $\alpha$-AgI and $\alpha$-CuBr we note that the metal ions have partial occupation and thus multiple possible positions. We therefore compute the total energies of all possible permutations of the positions of the mobile ions within the unit cell. The resulting minimum energy configurations are those in which the metal ions are located at the tetrahedral positions on adjacent faces, in agreement with AIMD calculations for $\alpha$-AgI~\citep{Wood2006}. Accordingly, to compute $U(r)$ for these materials we fix one of the tetrahedral metal ions and allow the other to move to all other permitted positions, as described above. 

Phonon calculations are performed using SG15 optimized norm-conserving Vanderbilt (ONCV) pseudopotentials~\citep{Hamann2013,Schlipf2015} and a PBE exchange-correlation functional, with a 60~Ry kinetic energy cutoff for wavefunctions. Sound velocities are derived from the phonon dispersions along the $\Gamma-X$ path for cubic cells, as defined in Ref.~\citenum{Setyawan2010}.

The volumetric images shown in Fig.~\ref{fig:potential_map} were generated in VESTA~\citep{Momma2011}.

\paragraph{\textit{Ab initio} molecular dynamics simulations --}
AIMD simulations are carried out using the SIESTA code~\citep{Soler2002}.
The forces are calculated using the local density approximation (LDA) of density functional theory~\citep{Ceperley1980}, and a Harris functional is used for the first step of the self-consistency cycle. 
The core electrons are represented by pseudopotentials of the Troullier-Martins scheme~\citep{Troullier1991}. 
The basis sets for the Kohn-Sham states are linear combinations of numerical atomic orbitals, of the polarized double-zeta type~\citep{Sanchez-Portal1997,Sanchez-Portal2001}.  
The $\Gamma$-point is used for Brillouin zone sampling. 
$\alpha$-AgI AIMD calculations are performed in 256 atom supercells. AIMD calculations for rocksalt structures are performed in 216 atom supercells. 
The temperature is controlled by means of a Nos\'{e} thermostat~\citep{Nose1984}. The integration time step used is 1~fs and the total integration time is 26~ps. The equilibration time varies between different temperatures and systems and is determined by examining the mean square displacement. 

\paragraph*{Calculations of the ionic mobility from existing experimental data --}
Experimental mobilities are found by fitting existing experimental data in the literature. We have assumed that there is only one mobile ion species. The conductivity is fitted using $\sigma=qn\mu$, where $n=N\exp\left(-E_{a}/k_BT\right)$ is the number of mobile ions. Here, $N$ is the total number of atoms of the mobile species in the crystal, $E_{a}$ is the activation energy necessary to make the ion mobile, $k_B$ is the Boltzmann constant and $T$ is the temperature. 

\section*{Acknowledgements}
We acknowledge the National Research Foundation, Prime Minister Office, Singapore, under its Medium Sized Centre Programme. A.R. thanks the support by Yale-NUS College (through Grant No. R-607-265- 380-121). The computational work was supported by the Centre of Advanced 2D Materials, funded by the National Research Foundation, Prime Minister's Office, Singapore, under its Medium-Sized Centre Programme.
\section*{Author Contributions}
A.R. and A.H.C.N. conceived the research. A.R. performed the theoretical derivations. K.N. performed the DFT calculations and analysis relating to the computation of the ion mobilities. A.C. performed the AIMD simulations and extracted all experimental mobilities. All authors contributed to the writing of the manuscript. 

%


\end{document}